\renewcommand\t{\tau}

\newcommand\D{\Delta}

\newcommand{\diracslash}[1]{#1\llap{/\kern2pt}}

\newcommand{\be}{\begin{equation}}
\newcommand{\ee}{\end{equation}}
\newcommand{\bea}{\begin{eqnarray}}
\newcommand{\eea}{\end{eqnarray}}
\newcommand{\ba}[1]{\begin{array}{#1}}
\newcommand{\ea}{\end{array}}

\documentclass[twocolumn,superscriptaddress,secnumarabic,amssymb,nobibnotes,nofootinbib,aps,pre,showpacs]{revtex4}
\usepackage{graphicx}
\begin{document}

\title{Wavelet analysis and scaling properties of time series}
\author{P. Manimaran}
 \affiliation{School of Physics, University of Hyderabad,
 Hyderabad 500 046, India}
\author{Prasanta K. Panigrahi}
\author{Jitendra C. Parikh}
\affiliation{ Physical Research Laboratory, Navrangpura, Ahmedabad
380 009, India}
\date{\today}

\def\be{\begin{equation}}
\def\ee{\end{equation}}
\def\bearr{\begin{eqnarray}}
\def\eearr{\end{eqnarray}}
\def\zbf#1{{\bf {#1}}}
\def\bfm#1{\mbox{\boldmath $#1$}}
\def\hf{\frac{1}{2}}

\begin{abstract}
We propose a wavelet based method for the characterization of the
scaling behavior of non-stationary time series. It makes use of
the built-in ability of the wavelets for capturing the trends in a
data set, in variable window sizes. Discrete wavelets from the
Daubechies family are used to illustrate the efficacy of this
procedure. After studying binomial multifractal time series with
the present and earlier approaches of detrending for comparison,
we analyze the time series of averaged spin density in the 2D
Ising model at the critical temperature, along with several
experimental data sets possessing multi-fractal behavior.
\end{abstract}

\pacs{05.45.Df, 05.45.Tp, 89.65.Gh}

\maketitle A large number of studies have been carried out to
analyze scaling properties of fluctuations in time series. The
studies have involved time series of dynamical variables from
physical, biological and financial systems \cite{mandel,feder}.
For scaling analysis different approaches have been suggested and
implemented, starting from the structure function method to
wavelet transform modulus maxima (WTMM) \cite{pen,arn1,phand} and
the recent, detrended fluctuation analysis (DFA)
\cite{ple,chen,khu} and its variants \cite{peng,matia,gopi}. The
difficulty in characterizing the scaling property stems from the
fact that, the observed time series is very often non-stationary.
Hence, it is essential to define fluctuations in a manner which
takes proper account of non-stationarity.

In this note, we propose a new method, based on discrete wavelet
transform \cite{daub}, to separate the trend in the time series
from the fluctuations. The method is direct and suggests itself
naturally from the basic concepts underlying wavelet
decomposition, apart from being supplementary to the detrended
fluctuation analysis. The fact that the so called low-pass
coefficients represent a coarse grained version of the data in
wavelet transform and the built-in ability of the wavelets to have
variable window sizes for coarse graining, makes it a natural tool
for identifying fluctuations around trends at various scales. We
use this method to examine scaling behavior of a time series. For
the purpose of checking the efficacy of our procedure and
comparison with multifractal detrended fluctuation analysis
(MF-DFA), we consider time series generated ($10^6$ data points)
from the binomial multifractal model \cite{feder}, for which the
scaling exponent is analytically calculable. The method has also
been checked on Gaussian random noise. We then analyzed the time
series of average spin densities ($9$x$10^5$ data points) in a
simulation of the 2D Ising model on a $256$ x $256$ lattice at the
critical temperature $T_c$, where each update of the system in the
simulation is taken as one time step \cite{hwa,osa}. These
computer generated time series are shown in Fig. 1. Experimentally
measured data \cite{jha} of ion saturation currents and floating
potential in Tokamak plasma are shown in Fig. 2. These are then
analyzed for their multi-fractal analysis.
\begin{figure}
\centering
\includegraphics[width=2.15in]{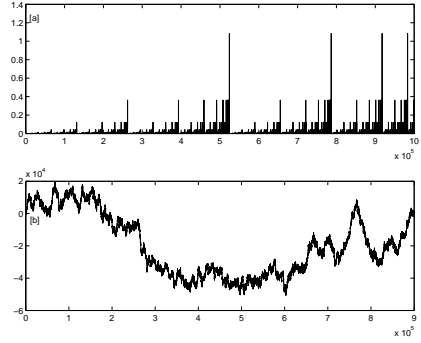}
\caption{Time series simulated through (a) binomial multifractal
model and (b) 2D Ising model at critical temperature.}
\end{figure}
\begin{figure}
\centering
\includegraphics[width=2.15in]{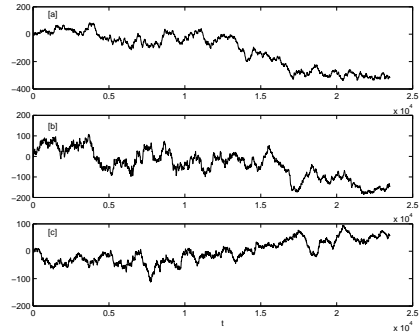}
 \caption{Time series of, (a)ion saturation current (IC),
(b) floating potential (FP), 6mm inside the main plasma and
 (c) ion saturation current (ISC), when the probe is in the limiter shadow.
 Each time series has approx. 24,000 data points.}
\end{figure}
Before we describe our approach in detail, it is worthwhile to
give a brief summary of some basic features of scaling. For this
purpose let $x(t_i)$ denote the value of an observable at
$t_i=i\Delta t$. The set ${x(t_i); i=1,2,..,N}$ is then the time
series under consideration. A simple way to define fluctuations,
for a stationary time series at time scale $\tau_k=k\Delta t$, is
    \be
\D x_i(\t_k) = x(t_i + \t_k) - x(t_i); i=1,2,..N-k.
    \ee
These fluctuations are said to have scaling property, if the
probability distribution function (pdf) of {$\Delta x_i(\tau_k)$}
has the same form for different values of $\tau_k$. Further, the
parameters that characterize the pdf depend in a well defined
manner on $\tau_k$. For example, for independent fluctuations with
a Gaussian pdf, we have the well-known scaling results for the
mean $ \epsilon(k \Delta t)= k \epsilon(\Delta t) $ and variance
$\sigma^2 (k \Delta t) = k \sigma^2 (\Delta t)$. In general for
all finite moments $m_q$ of order $q$, this is expressed as,
    \be
m_q = \langle|(\Delta x(k\Delta t))|^q\rangle = \frac{1}{N-k}
\sum_{i=1}^{N-k} |\Delta x_i(k\Delta t)|^q \sim (k \Delta
t)^{\zeta(q)}
    \ee
where $\zeta(q)$ is constant for mono-fractals. The Hurst exponent
$H=\zeta(q=2)$, equals $0.5$ for the Gaussian white noise. The
scaling property of the pdf considered above implies that, the
variability at different time scales in time series is fractal in
nature; it is self-similar (more precisely self-affine).
Mono-fractals with $H < 0.5$ are long range anti-correlated and
for $H > 0.5$ the signal shows long range correlation. For time
series of complex systems, it often turns out that scaling is
present but the dependence of scale factor on q is not linear but
decreases with increasing $q$. This type of scaling behavior is
termed as multi-fractal.

The scaling (multi-scaling) property of the time series arises
from the corresponding property of the fluctuations. More
precisely, in this connection, the probability distribution and
the time correlations of the fluctuations are the properties of
importance. There are complications and difficulties, if
conclusions about the scaling behavior of a dynamical system, are
entirely based on a finite length time series in presence of
correlations. This is the case for most physiological and
financial time series and the associated problems have been
highlighted in \cite{taqqu,bouch,xu}.

The difficulties of standard multi-fractal formalism, based on
structure function method, led to the development of WTMM, a
continuous wavelet transform based approach. Although like the
structure function method, this is also a global approach, the
built-in advantage of wavelets in identifying scaling properties
of data, made WTMM quite ideal in finding the multi-fractal
behavior. In later stage, the multi-fractal detrended fluctuation
analysis (MF-DFA) has been developed to tackle the non-stationary
time series. The basic idea in the above approach is to isolate
fluctuations in the data set, through multiple local windows, of
varying sizes. For this purpose, once the window size is fixed, an
appropriate fit e.g., a polynomial fit, is employed to identify
the trend and the fluctuations are then isolated, by subtracting
the trend from the data points.  The method we propose is based on
the fact that, the low-pass coefficients in wavelet transform,
resemble the data, albeit in an averaged manner. The extent of
averaging depends on the level of wavelet decomposition. It
supplements the MF-DFA, in the sense that, instead of a polynomial
fit, one can use the appropriate low-pass coefficients for
capturing the trend.

The basis elements in discrete wavelet transform, provide a
complete and orthonormal set, unlike the continuous wavelets,
wherein the basis functions usually comprise an  over complete
set. The two key members are the scaling (or father wavelet)
$\phi(t)$ and the mother wavelet $\psi(t)$, respectively
satisfying $\int dt \phi(t)=A$ and $\int dt \psi(t)=0$. Here $A$
is a constant; $\phi(t)$ and $\psi(t)$ satisfy square
integrability conditions apart from being orthogonal to each
other. The two key operations, underlying the construction of a
complete orthonormal basis set, are translation and scaling of the
father and mother wavelets, which have strictly finite sizes.
Translation by discrete steps brings in the index $k$ to both
father and mother wavelets: $\phi_k(t)\equiv \phi(t-k)$ and
$\psi_k(t)$. Producing daughter wavelets, copies of the mother
wavelet, albeit thinner and taller, through scaling allows one to
form a complete set. With the scaling index $j$, conveniently
running from $0$ to $\infty$, wavelets can be compactly written as
$\psi_{j,k}(t)$, where $\psi_{0,0}(t)$ is the original mother
wavelet. The key equation underlying all wavelets is the
multi-resolution-analysis (MRA) equation:
\bearr
\phi(t)=\sum_n h(n) \sqrt(2) \phi(2t-n) \nonumber \\
\psi(t)=\sum_n \tilde{h}(n) \sqrt(2) \phi(2t-n). \eearr Here,
$h(n)$ and $\tilde{h}(n)$ are the low and high-pass filter
coefficients satisfying the constraints, $\sum_n h(n)= \sqrt{2}$,
$\sum_n h(n)h(n-2k)=\delta_{k,0}$, $\sum_n \tilde{h}(n)=0$,
$\sum_n \tilde{h}(n) \tilde{h}(n-2k)=\delta{k,0}$ and $\sum_n
\tilde{h}(n) h(n-2k)=0$ originating from the normalization and
orthogonality conditions mentioned above. Here $n$ is the length
of the filter coefficients. The MRA equation can be used to obtain
recurrence relations
   \bearr
c_j(k)=\sum_n h(n-2k) c_{j+1}(n) \nonumber\\
d_j(k)=\sum_n \tilde{h}(n-2k) c_{j+1}(n).
  \eearr
Here, $c_j(k)$ and $d_j(k)$ are respectively the low-pass and
high-pass coefficients at level $j$; MRA equation implies that,
both of these coefficients can be obtained from the next level
low-pass coefficients alone. We note that, the low-pass
coefficient is given by, $ c_{j+1}(k)=\int dt \phi_{j+1,k}(t)
f(t)$, where $f(t)$ is the function or data under consideration.
In the limit, $j\rightarrow\infty$, the scaling function tends to
the Dirac delta function; hence the corresponding low-pass
coefficient $c_{j\rightarrow\infty}(k)$ is the value of the
function at location $k$. Therefore, starting from the values of
the function at the highest resolution, one can find all the
scaling and high-pass coefficients, without explicitly knowing the
wavelet basis elements.

In a broad sense, the low-pass  coefficients capture the trend and
the high-pass coefficients keep track of the fluctuations in the
data. In case of the simplest Haar wavelet ($n=2$),
$h(0)=h(1)=\frac{1}{\sqrt{2}}$ and
$\tilde{h}(0)=-\tilde{h}(1)=\frac{1}{\sqrt{2}}$. The low-pass and
high-pass or wavelet coefficients are respectively the averages
and differences of data points. In case of other discrete
wavelets, these coefficients are appropriately weighted averages
and differences. For example, Daubechies-4 (Db-4) wavelet is
characterized by four filter coefficients. It is worth mentioning
that, the Daubechies family of wavelets are made to satisfy
vanishing moments conditions:$\int dt~ t^m\psi_{j, k}(t)=0$. This
makes them blind to variations in a data set, which can be
captured by polynomials of suitable degree, thereby making them
ideal for separating fluctuations from average behavior. This
property will be made use of extensively in the present
manuscript. Wavelets are naturally endowed with an appropriate
window size, which manifests in the scale index or level, and
hence can capture the local averages and differences, in a window
of one's choice.

The data for wavelet analysis is assumed to be of the size $2^L$,
in case the same is not available recourse is taken to padding. In
the present analysis, we have used constant padding at the ends.
For the above data, one can have a maximum $L$ level
decomposition, although one can stop at any level below $L$. The
level-1 high-pass coefficients are half the size of the data and
represent fluctuations at the lowest scale. The progressively
higher level coefficients represent fluctuations at larger scales;
the low-pass coefficients at a given level represent the
appropriately averaged data, commensurate with the window size of
the level concerned. It should be mentioned that, for wavelets
other than Haar, artifacts, both in low and high-pass
coefficients, arise at the end points, due to the need for
circular or other forms of required extensions. We have used
constant padding at both the ends, discarding the coefficients
originating from them, to avoid the edge effect because of
wrapping of wavelet filters by the convolution process during
decomposition.

In the present approach, a time series or the cumulative
distribution in case of binomial multifractal data have been
subjected to a multi-level decomposition. Reconstructed series
after removal of successive high-pass coefficients was subtracted
from the data to extract the fluctuations, $F(s)$ as a function of
scale $s$. Here, $s$ denotes the level of decomposition. As is
clear, reconstruction after removing the first level high-pass
coefficients achieves coarse graining in a window size, which
depends on the length of the filter coefficients. For a given
wavelet, removal of other level high-pass coefficients enlarges
the window size. As has been mentioned in the beginning, the fact
that in Daubechies (Db) family of wavelets, the low-pass captures
the appropriate polynomial behavior of the data set, makes them
quite useful for our analysis. We have used Db-4 to Db-20 basis
sets for comparison. It was observed, for the data under
consideration here, that after a certain point, the improvements
in the scaling exponents due to a higher wavelet was minimal.
\begin{figure}
\centering
\includegraphics[width=2.15in]{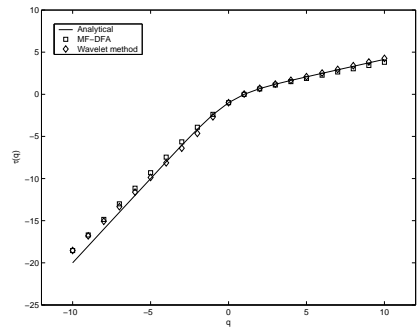}
 \caption{Wavelet method ($\diamondsuit$) and MF-DFA ($\Box$) analysis of computer
 generated time series for binomial multifractal model show the dependence of $\t(q)$ on
 $q$, which compares quite well with the analytically calculated $\t(q)$ values
(dashed line). For $q < 0$, Db-4 and for $q > 0$, Db-14 wavelets
have been used}
\end{figure}
Analogous to detrended fluctuation analysis, one can calculate the
scaling exponents from the behavior of $F_q(s)$ defined as,
    \be
 F_q(s)\equiv \{\frac{1}{N}\sum_{k=1}^N |F(s)|^q\}^{1/q}.
    \ee
Here $q$ is can take both positive and negative integral values,
except zero.  If the time series, under analysis, possesses
fractal behavior, then  $F_q(s)$ reveals a power-law scaling:
    \be
F_q(s) \sim s^{h(q)}.
    \ee
As noted earlier, if $H$ is constant for all $q$ then the series
is monofractal, otherwise it is multifractal. For $q<0$, $h(q)$
captures the scaling properties of the small fluctuations, whereas
for $q>0$ those of the large fluctuations. Often, the scaling
exponent $h(q)$ is represented in terms of $\t(q)$, where $\t(q) =
q h(q) - 1$.
\begin{figure}
\centering
\includegraphics[width=2.15in]{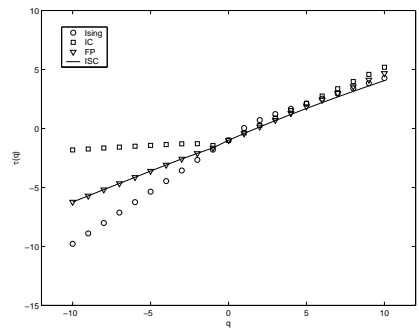}
\caption{Wavelet based fluctuation analysis of the experimental
data (shown in Fig. 2)and spin density values of the 2D Ising
model at $T_c$, shows the dependence of $\t(q)$ on various values
of $q$.}
\end{figure}
In Table-1, the $h(q)$ values are given for the binomial
multi-fractal time series, using analytical method, MF-DFA and the
present discrete wavelet based approach. In MF-DFA, we have used a
quadratic polynomial fit, to isolate the trend. A host of wavelets
have been tested, the results shown here correspond to Db-14 for
$q > 0$ and Db-4 for $q < 0$, since the improvement was minimal
after that.
\begin{table}
\centering
\begin{tabular}{ccccc}
\hline \hline
  q & $h(q)_{BMFS_a}$ & $h(q)_{BMFS_s}$ & $h(q)_{BMFS_w}$ \\
  \hline \hline

  -10.0000    &1.9000    &1.7544    &1.7534\\
   -9.0000    &1.8889    &1.7439    &1.7558\\
   -8.0000    &1.8750    &1.7307    &1.7588\\
   -7.0000    &1.8572    &1.7138    &1.7627\\
   -6.0000    &1.8337    &1.6914    &1.7677\\
   -5.0000    &1.8012    &1.6605    &1.7746\\
   -4.0000    &1.7544    &1.6154    &1.7845\\
   -3.0000    &1.6842    &1.5470    &1.7998\\
   -2.0000    &1.5760    &1.4551    &1.8224\\
   -1.0000    &1.4150    &1.3802    &1.6806\\
         0    &     0    &     0    &     0\\
    1.0000    &1.0000    &0.9923    &1.0316\\
    2.0000    &0.8390    &0.8169    &0.8538\\
    3.0000    &0.7309    &0.6995    &0.7384\\
    4.0000    &0.6606    &0.6253    &0.6660\\
    5.0000    &0.6139    &0.5771    &0.6193\\
    6.0000    &0.5814    &0.5444    &0.5878\\
    7.0000    &0.5578    &0.5212    &0.5655\\
    8.0000    &0.5400    &0.5040    &0.5490\\
    9.0000    &0.5261    &0.4907    &0.5365\\
   10.0000    &0.5150    &0.4803    &0.5267\\
  \hline \hline
\end{tabular}
\caption{The h(q) values of binomial multi-fractal series (BMFS)
computed analytically ($BMFS_a$), through MF-DFA ($BMFS_s$) and
wavelet ($BMFS_w$) approach. For $q < 0$, Db-4 and for $q
> 0$, Db-14 wavelets have been used.}
\end{table}

As is clear from the table, for positive $q$, the wavelet estimate
of the Hurst exponent for the binomial multi-fractal series is
extremely reliable when compared with the analytical result. For
$q<0$, when Db-14 was used in the wavelet based approach, smaller
$h(q)$ values were obtained. However, $h(q)$ values decreased with
increasing values of $q$. It was found that, the results for $q<0$
improves substantially if one uses a lower order Daubechies
wavelet, e.g., Db-4. It should be noted that, this amounts to
capturing the trend by a lower order polynomial curve, like in the
MF-DFA approach. Higher order wavelets, having a large number of
filter coefficients average the data over a much bigger window
size and hence are not expected to perform well in estimating the
smaller fluctuations. Hence, in our analysis a higher order
wavelet has been used for $q>0$ and a lower order one for $q<0$.
These results are shown in Fig. 3. Analysis of the scaling
properties of the spin density fluctuations and experimental data
sets reveal that the corresponding time series have multi-scaling
behavior, as is seen clearly in Fig. 4.

In conclusion, the wavelet based method presented here, for
calculating the scaling exponents, is found to be quite efficient,
fast in computation and reliable. It performed well for both
synthesized and experimental data. Our method is well suited to
characterize both large ($q>0$) and small ($q<0$) fluctuations. In
the later case, one needs to use a lower order wavelet, since the
larger size of the filter coefficients of the higher order
wavelets average over a bigger window size, thereby distorting the
smaller fluctuations. As compared to MF-DFA, the discrete wavelet
based approach has less number of windows. This procedure
compliments the former in the sense that, fluctuations at
different scales have been isolated by subtracting the local
polynomial trend captured through Daubechies family of wavelets.

The authors are grateful to Dr. R. Jha for providing the
experimental data for our analysis.

\end{document}